\documentclass{aa} 

\usepackage{siunitx}
\usepackage{graphicx}
\usepackage{txfonts}
\usepackage{hyperref}

\begin{document}

   \title{The impact of numerical oversteepening on the fragmentation boundary in self-gravitating disks}
   \titlerunning{Oversteepening in self-gravitating disks}

   \subtitle{}

   \author{J. Klee \inst{1}\fnmsep\thanks{jklee@astrophysik.uni-kiel.de}
          \and
          T. F. Illenseer\inst{1}\fnmsep\thanks{tillense@astrophysik.uni-kiel.de}
          \and
          M. Jung\inst{2}\fnmsep\thanks{manuel.jung@hs.uni-hamburg.de}
          \and
          W. J. Duschl\inst{1,3}\fnmsep\thanks{wjd@astrophysik.uni-kiel.de}
		  }

   \institute{Institut für Theoretische Physik und Astrophysik, Christian-Albrechts-Universität zu Kiel,
              Leibnizstr. 15, 24118 Kiel, Germany \\
              \and 
          	  Hamburger Sternwarte, Universität Hamburg, 
              Gojenbergsweg 112, 21029 Hamburg, Germany \\
              \and
              Steward Observatory, The University of Arizona,
              933 N. Cherry Ave., Tucson, AZ 85721, USA \\
              }

   \date{Accepted July 8, 2017}


  \abstract
   {It is still an open issue whether a self-gravitating accretion disk fragments. There are many different physical and 
   numerical explanations for fragmentation, but simulations often show a non-convergent behavior for ever better 
   resolution.}
   {We investigate the influence of different numerical limiters in Godunov type schemes on the fragmentation boundary 
   in self-gravitating disks.}
   {We compare the linear and non-linear outcome in two-dimensional shearingsheet simulations using the 
   \texttt{VANLEER} and 
   the \texttt{SUPERBEE} limiter.}
   {We show that choosing inappropriate limiting functions to handle shock-capturing in Godunov type schemes can lead 
   to an overestimation of the surface density in regions with shallow density gradients. The effect amplifies itself on
   timescales comparable to the dynamical timescale even at high resolutions. This is exactly the environment, where 
   clumps are expected to form. The effect is present without, but scaled up by, self-gravity and also does not 
   depend on cooling. Moreover it can be backtracked to a well known effect called oversteepening. If the effect is 
   also observed in the linear case, the fragmentation limit is shifted to larger values of the critical cooling 
   timescale.}
   {}


   \maketitle
%

\section{Introduction}
If a disk becomes massive enough, compared to its enclosed central object, self-gravity plays a major role in the
further evolution of the disk. This is of relevance in protoplanetary and AGN-disks, at least in the 
early phase. The resulting gravitational instabilities can either drive angular momentum transport by settling into a 
gravito-turbulent state \citep{paczynski_model_1978}, or lead to fragmentation and thus yield a possible mechanism 
for planet and star formation \citep{boss_astrometric_1998,levin_stellar_2003}. A disk becomes gravitationally 
unstable if the Toomre-parameter
\begin{equation}
    Q = \frac{c_{s} \kappa}{\pi G \Sigma}
\end{equation}
fulfills $Q  \lesssim 1$  \citep{toomre_gravitational_1964}. Here $c_{\mathrm{s}}$ is the speed of sound, 
$\kappa$ the epicyclic frequency, $\Sigma$ the surface density and $G$ the gravitational constant. 
Throughout this paper we will assume that the angular velocity $\Omega$ is close to its Keplerian value. In 
that case $\kappa = \Omega$ holds.
   
The exact conditions which lead to fragmentation is subject of debate. \citet{gammie_nonlinear_2001} 
showed that if the cooling timescale $\tau_{\mathrm{c}}$ falls short of the dynamical 
timescale $\Omega^{-1}$, i.e., the dimensionless relative cooling time scale
\begin{equation}
   \beta = \tau_{\mathrm{c}} \Omega \lesssim 3,
\end{equation}
fragmentation occurs. More recent publications, however, show that with better resolutions, fragmentation can also 
occur for larger values of $\beta_{\mathrm{crit}}$, i.e., slower cooling 
\citep{meru_non-convergence_2011,meru_convergence_2012}. This 
non-convergence of numerical simulations remains an ongoing problem. \citet{rice_investigating_2005} even stress that 
$\beta_{\mathrm{crit}}$ should be used at all for investigating the onset of fragmentation. Instead, the maximum stress 
that a disk can carry before it is fragmenting should be used. This is represented by a maximum value for the 
dimensionless parameter $\alpha$ \citep{shakura_black_1973}.  Furthermore, it is not clear whether numerical or 
physical effects are responsible for this fragmentation. \cite{cossins_characterizing_2009} check for a wide 
range of parameters in order to characterize self-gravitating disks in a more encompassing manner. The current state of 
the debate was reviewed recently by \cite{kratter_gravitational_2016-2} and 
\cite{rice_evolution_2016}.

On the numerical side, there are several effects which can be present in different kind of codes. 
\cite{lodato_resolution_2011} take artificial viscosity into account as a possible explanation for 
fragmentation. This is investigated further by \cite{meru_convergence_2012}. They tune their parameters defining the 
artificial viscosity in SPH and grid-based simulations (\texttt{Fargo}, see \citealt{masset_fargo:_2000}) in a way to 
maximize $\beta_{\mathrm{crit}}$, thus minimizing artificial viscosity, which goes along with a reduction of numerical 
heating. They also state that the tuned parameters are not in the ideal regime against particle penetration. Their 
results lead to large critical relative cooling timescales $\beta_{\mathrm{crit}} \lesssim 30$, at which fragmentation 
occurs, which differ from results for grid-based and other SPH simulations. 
\cite{rice_convergence_2012,rice_convergence_2014} argue that this 
procedure causes unforeseeable numerical errors, and state that the problem is more about the cooling function, which 
is not properly introduced within the SPH codes, leading to more likely fragmentation at higher resolutions. With this 
they reach a fragmentation boundary at around $6 \lesssim \beta_{\mathrm{crit}} \lesssim 8$. 
\cite{muller_treating_2012} stress that in two-dimensional simulations gravity needs to be smoothed in order to avoid 
singular behavior at higher resolutions. Simulations by \cite{young_dependence_2015} show indeed that their results 
change fundamentally when smoothing out the gravity to Keplerian vertical pressure scale height $H$, because of the  
neglected vertical structure. They state that within their framework, where two possible fragmentation modes 
are present, the second one, driven by quasi-static collapse, shows up only if the third dimension is taken into 
account. This approach needs to resolve gravitational forces on scales below the vertical scale-height $H$, which is 
not possible in 2D, because of the above mentioned limitations. In order to investigate non-quasi-static fragmentation 
a two-dimensional model is, however, still valid and has been used, for instance, by the same authors later
\citep{young_quantification_2016}. Another numerical effect may arise by additional edge effects, 
emerging from the boundaries in global simulations \citep{paardekooper_numerical_2011} and leading to fragmentation at 
higher values of $\beta$. They solve this problem by adjusting $\beta$ from a very large value slowly to the actual 
one. 

On the physical side \cite{hopkins_turbulent_2013} show that a turbulent disk is never 
stable. They use a stochastic approach and find that there is always a probability of a turbulent,
self-gravitating disk to fragment. \cite{rice_evolution_2016}, however, claims that this is due to the isothermality of the disks and 
to the feedback cycle of gravitational instabilities is not being treated appropriately. Similar assumptions let 
\cite{lin_gravitational_2016} revise the Toomre criterion itself, which actually applies only to a 
laminar, 
inviscid disk in which 
there are no net thermal losses. They conclude that adding cooling and viscosity leads to secondary instabilities which 
can drive fragmentation and that there formally exists no region for $\beta$ or the maximum viscous stress $\alpha$ 
where the disk is stable. That there are two modes of fragmentation, one for $\beta_{\mathrm{crit}} \lesssim 3$ at free 
fall timescale 
and one for $\beta_{\mathrm{crit}} \lesssim 12$ with a quasi-static collapse, is mentioned by 
\cite{young_dependence_2015}. For the 
second mode, $\beta_{\mathrm{crit}}$ is derived by comparing cooling timescales with the timescales of encountering a 
spiral wave 
assuming the tightly-wound approximation. \cite{young_dependence_2015} conclude that two-dimensional calculations are not 
suitable to simulate the quasi-static part of the fragmentation, because these require gravitational smoothing, which 
suppresses a quasi-static collapse at 
the same time.

Moreover \cite{paardekooper_numerical_2012} shows in the shearingsheet that fragmentation has a stochastic nature. He 
runs simulations with the same setup multiple times, which differ only in the subsonic random velocity field. 
\cite{young_quantification_2016} quantify this effect and come to the conclusion that stochastic fragmentation 
can, in general, not change the radii of fragmentation significantly.

Outline: In section~\ref{sec:methods} we will explain the underlying numerics. In section~\ref{sec:testing} we will show 
tests of the code, making extensive use of the linear theory test used by \citet{gammie_nonlinear_2001} and 
\citet{paardekooper_numerical_2012}. The nonlinear outcome will be presented in section~\ref{sec:results} where the 
value of $\beta_{\mathrm{crit}}$ and the stochastic nature is examined. In section~\ref{sec:discussion} further 
numerical effects 
and their impacts on the results are discussed as well as the expansion to other kinds of codes. In 
section~\ref{sec:conclusion} we summarize and discuss our results.
    
\section{Methods}\label{sec:methods}
We use the hydrodynamic simulation suite \texttt{Fosite}\footnote{http://www.astrophysik.uni-kiel.de/fosite/} 
\citep{illenseer_two-dimensional_2009} which is based on an enhanced method of a Godunov type scheme 
\citep{kurganov_new_2000}. Generally, it can calculate on arbitrary two-dimensional curvilinear, orthogonal grids and 
is easily expandable, because of its modular design. 

The shearingsheet-approximation is implemented within this framework, i.e., modules for fictitious forces, sheared 
boundary conditions, a Poisson-solver and the timescale-parameterized cooling by \cite{gammie_nonlinear_2001} are 
implemented.

\subsection{System of Equations}
Similarly to \citet{hawley_local_1995} and \citet{gammie_nonlinear_2001} we consider a small comoving region within a 
flat disk in plane-polar geometry at a fixed radius $r=r_0$ and the angle $\varphi = \varphi_0+\Omega(r_0) t$, where 
$t$ is the time and $\Omega(r_0) = \Omega_0$ the angular velocity at $r_0$. For abbreviation we drop the index of 
$\Omega_0$ in the following. We introduce local coordinates $x=r - r_0$ and $y = r_0(\varphi - \Omega t)$ and expand 
the equations of motion to first order in $|x|/r_0$ \citep{goldreich_ii._1965}. This yields the following system of 
equations:
\begin{subequations}
\label{eq:navier_stokes}
\begin{align}
    \frac{\partial \Sigma}{\partial t} + \nabla \cdot \left( \Sigma \mathbf{v} \right) = & 0 
    \label{eq:navier_stokes_a}\\
    \frac{\partial \Sigma \mathbf{v}}{\partial t} + \nabla \cdot \left(\Sigma \mathbf{v}\otimes \mathbf{v} + P 
    \mathbb{I} \right) = & \Sigma \left(-2\mathbf{\Omega} \times \mathbf{v} + 2 q \Omega^2 x \mathbf{\hat{e}_x} 
    \right)  \nonumber \\
                         & - \Sigma \nabla \Phi_{\mathrm{sg}} \label{eq:navier_stokes_b}\\  
    \frac{\partial E}{\partial t} + \nabla \cdot \left( \left(E + P \right) \mathbf{v} \right) = 
    & \Sigma \left(-2 \mathbf{\Omega} \times \mathbf{v} + 2 q \Omega^2 x \mathbf{\hat{e}_x}  \right) \cdot \mathbf{v} 
    \nonumber \\
    & -\Sigma \nabla \Phi_{\mathrm{sg}} \cdot \mathbf{v}+\Sigma Q_{\mathrm{cool}} \label{eq:navier_stokes_c}
\end{align} 
\end{subequations}
Here, $\mathbf{v}$ is the velocity and $E$ the total energy density. Additionally, we 
have $P$ the vertically integrated pressure, $\mathbf{\Omega} = \Omega \mathbf{\hat{e}}_z $ the oriented angular 
velocity and $q = -\left. \frac{\mathrm{d} \ln{\Omega}}{\mathrm{d} \ln{r}} \right|_{r_0}$ the shearing parameter with 
$q=3/2$ in the 
keplerian case. $q=3/2$ is used throughout this paper. $\Phi_{\mathrm{sg}}$ is the potential induced by self-gravity in 
the midplane of the disk. It connects to the system via Poisson's equation
\begin{equation}\label{eq:poisson}
\Delta \Phi_{\mathrm{sg}} =   4 \pi G \Sigma \delta(z),
\end{equation}
where $\delta(z)$ is the vertical density structure in the $z$-direction. The parametrized cooling is
\begin{equation}\label{eq:cooling}
     Q_{\mathrm{cool}} = \frac{p \Omega}{(\gamma -1)\beta}
\end{equation}
with $\gamma$ the two-dimensional heat capacity ratio. 
 
In order to close the system of equations, we assume an ideal gas equation of state 
\begin{equation}
    p = (\gamma - 1) \Sigma \epsilon,
\end{equation}
with the pressure $p$ and the specific internal energy $\epsilon$. The latter is related to the total energy density according to
\begin{equation}
    E = \tfrac{1}{2} \Sigma \mathbf{v}^2 + \Sigma \epsilon.
\end{equation} 

In order to allow for larger timesteps we split the constant velocity offset by introducing the fiducial velocities 
$\mathbf{v}' = \mathbf{v} + q\Omega x \mathbf{\hat{e}}_y$, with $v_x' = v_x$ and $v_y' = v_y + q \Omega x 
\mathbf{\hat{e}}_y$ \citep{masset_fargo:_2000,mignone_conservative_2012}.This yields the new system 
of equations:
\begin{subequations}
\label{eq:navier_stokes_fargo}
\begin{align}
    \frac{\partial \Sigma}{\partial t} + \nabla \cdot \left( \Sigma \mathbf{v'} \right) - q\Omega x \frac{\partial 
    \Sigma}{\partial y} 
    = 0 \label{eq:navier_stokes_fargo_a}\\
    \frac{\partial \Sigma \mathbf{v'}}{\partial t} + \nabla \cdot \left(\Sigma 
    \mathbf{v'}\otimes \mathbf{v'} + p \mathbb{I} \right) - q \Omega x \frac{\partial \Sigma \mathbf{v}'}{\partial y}=  
    & \nonumber\\ 
    \Omega \Sigma \begin{pmatrix} 2 v_y' \\ (q-2) v_x' \end{pmatrix}  
    & - \Sigma \nabla \Phi_{\mathrm{sg}} \label{eq:navier_stokes_fargo_b}\\
    \frac{\partial E'}{\partial t} + \nabla \cdot \left( \left(E' + p \right) \mathbf{v'} \right) - q\Omega x 
    \frac{\partial E'}{\partial y}= \nonumber \\
    q \Omega \Sigma v_x' v_y' - \Sigma \nabla &\Phi_{\mathrm{sg}} \cdot \mathbf{v'}+ \Sigma Q_{\mathrm{cool}}, 
    \label{eq:navier_stokes_fargo_c}
\end{align}
\end{subequations}
with the reduced energy density
\begin{equation}
    E' = \tfrac{1}{2} \Sigma \mathbf{v'}^2 + \Sigma \epsilon.
\end{equation}
The Fargo scheme \citep{masset_fargo:_2000} was already implemented in \texttt{Fosite} 
\citep{jung_multiskalensimulationen_2016} and can be enabled for $\mathbf{v} = \mathbf{v'} + \mathbf{w}$, 
where 
$\mathbf{w}$ is an arbitrary solenoidal velocity \citep{mignone_conservative_2012}.

\subsection{Fictitious Forces}
The source terms for fictitious forces are implemented in two different ways according to the right hand side of 
eqs.~\ref{eq:navier_stokes_b},~\ref{eq:navier_stokes_c} and 
eqs.~\ref{eq:navier_stokes_fargo_b},~\ref{eq:navier_stokes_fargo_c}, respectively. The implementation differs depending 
on whether advection splitting is enabled or not.

\subsection{Boundary Conditions}
We chose periodic boundary conditions in $y$-direction and sheared periodic boundary conditions in $x$-direction 
\citep{hawley_local_1995}. With $f \in \{\Sigma, v_{x}, v_{y}', p \}$ and $L_x$, $L_y$ the size of the field in both 
directions we have
\begin{equation}
    f\left(x,y \right) = f\left(x, y+L_y \right)
\end{equation}
in $y$-direction and
\begin{equation}\label{eq:boundaries}
    f \left(x,y \right) =  f\left(x + L_x, y-q\Omega L_x t \right).
\end{equation}
in $x$-direction. If necessary, the data is linearly interpolated between adjacent cells \citep{hawley_local_1995}.

\subsection{Self-gravity}
We consider a razor thin disk, which means that $\delta(z)$ in eq.~\ref{eq:poisson} is the $\delta$-distribution, and 
solve this equation via common practice FFT routines\footnote{We use the following libraries: \texttt{FFTW} 
\citep{frigo_fftw:_2012} on scalar, \texttt{fft} (from MathKeisan package, see \url{www.mathkeisan.com}) on 
NEC-SX systems.} similar to the way \cite{gammie_nonlinear_2001} did. We shift the surface density field to the next 
periodic point, transform the result 
and calculate components of the 
potential via
\begin{equation}\label{eq:poisson_component}
    \Phi_{\mathrm{sg}} (\mathbf{k}(t)) = 2\pi G \frac{\Sigma(\mathbf{k}(t))}{|\mathbf{k}(t)|}.
\end{equation}
Again transforming and shifting back gives us the solution for $\Phi_{\mathrm{sg}}(x,y)$. The gravitational 
acceleration is obtained computing second order finite differences. We thereby also cut of all wavenumbers 
$\mathbf{|k|} = \pi / \left(\sqrt{2} \Delta x\right)$ \citep{gammie_nonlinear_2001}. \cite{young_dependence_2015} argue 
that this mimics the effect of a gravitational softening with length scale $\sim \SI{0.3}{H}$, where \si{H} is the 
scale height of the disk.

\subsection{Viscosity}
In order to measure the viscous stresses in a disk, the $\alpha$ prescription can be used 
\citep{shakura_black_1973}. Taking into account the cooling function in eq.~\ref{eq:cooling} an analytical expression 
for the dimensionless parameter $\alpha$ can be found \citep{gammie_nonlinear_2001}:
\begin{equation}\label{eq:alpha_theoretical}
    \alpha = \frac{1}{q^2 \gamma (\gamma - 1) \beta}.
\end{equation}
Simultaneously the $\alpha$-parameter can be calculated from the simulations assuming a Jacobi-like integral 
\citep{gammie_nonlinear_2001}, by splitting the stress-tensor $T_{xy}$ into a hydrodynamical part
\begin{equation}
    \langle H_{xy} \rangle = \langle \Sigma v_x' v_y' \rangle,
\end{equation}
with the residual velocities $v_x'$, $v_y'$ and a gravitational part
\begin{equation}
    \langle G_{xy} \rangle = \sum_{\mathbf{k}} \frac{\pi G k_x k_y}{|\mathbf{k}|^3} \left|\Sigma(\mathbf{k}) \right|^2.
\end{equation}
Together these yield
\begin{equation}\label{eq:alpha_numerical}
    \alpha = \frac{\langle H_{xy} \rangle + \langle G_{xy} \rangle}{q \langle \Sigma c_{\mathrm{s}}^2 \rangle}.
\end{equation}
Thus we can compare the numerical solution eq.~\ref{eq:alpha_numerical} with the analytical prediction in
eq.~\ref{eq:alpha_theoretical}.

\subsection{Numerical Viscsosity}
In order to resolve shocks, numerical schemes need to introduce non-linear terms. In SPH or finite difference schemes 
this is, for example, achieved by artificial viscosity. \texttt{Fosite} is based on finite volume methods
which introduce limiters in order to handle the shock regions appropriately. Otherwise artificial numerical 
oscillations will occur in the vicinity of discontinuities. According to Godunov's theorem 
\citep{godunov_difference_1959} this is a fundamental limitation of linear difference schemes with higher than first 
order spatial accuracy. Essentially, the limiter is switching to first-order accuracy in the shock regions. In smooth 
regions an ideal limiter should not affect the numerical scheme. In the real world, however, any limiter 
effects the solutions even though the slopes are moderate. Slope-limiters, like in our 
case, are applied to the slopes of the physical quantities itself whereas flux-limiters, which were used in 
\cite{paardekooper_numerical_2012}, are applied to the fluxes. With a slope $\Delta$, we introduce
\begin{equation}
    \bar{\Delta} = \phi(s) \Delta,
\end{equation}
where $\phi(s)$ is the limiter which depends on the ratio of neighboring slopes $s$.

Within this text, we use two different limiters, \texttt{VANLEER} \citep{van_leer_towards_1974} and \texttt{SUPERBEE} 
\citep{roe_algorithms_1982}. \texttt{VANLEER} tends to lead to more numerical dissipation and resolves shocks less good 
than \texttt{SUPERBEE}. The latter one, however, can oversteepen the slopes in shallow regions. A more thoroughly 
discussion on this issue can be found in \cite{toro_riemann_2009}, who gives an overview of some limiters applied to 
linear advection. This goes conform with the guidelines of \citet[ch. 8.3.4]{hirsch_numerical_2007}, who 
additionally states that 'limiters with continuous slopes generally favor convergence'. Whereas \texttt{VANLEER} has a 
continuous slope, limiters like \texttt{SUPERBEE} or \texttt{MINMOD} \citep{roe_characteristic-based_1986} show 
discontinuities in their slopes. The 
\texttt{VANLEER} limiter is described by
\begin{equation}
    \phi_{\mathrm{vl}}(r) = \frac{r + |r|}{1+|r|},
\end{equation}
and \texttt{SUPERBEE} by
\begin{equation}
 \phi_{\mathrm{sb}}(r) =
    \begin{cases} 
       0 & r\leq 0 \\
       \max \left(0,\min \left(2|r|,1\right),\min \left(|r|,2\right)\right) & r>0
    \end{cases}.
\end{equation}

\section{Tests \& Verification}\label{sec:testing}
We test our code with epicyclic motion \citep{stone_implementation_2010,paardekooper_numerical_2012} and linear 
theory \citep{gammie_nonlinear_2001,paardekooper_numerical_2012}. Thereby the linear theory test is being used 
extensively in order to obtain insights into the deviations between linear and nonlinear simulations.

\subsection{Epicyclic Motion}
The epicyclic motion test is adapted from \cite{paardekooper_numerical_2012}. We use exactly the same initial 
conditions with a resolution of $N_x = N_y = 128$, a field size of $L_x = L_y = 1$, an initial velocity perturbation 
$v_{x,0} = 0.1 c_{\mathrm{s}}$ and $v_{y,0} = 0$, with a speed of sound $c_{\mathrm{s}}= 0.01$. Then, we have 
oscillating solutions in the absence of any gradients
\begin{align}
    v_{x}(t) &= v_{x,0} \cos{\Omega t} + 2 v_{y}
    \sin{\Omega t} \\
    v_{y}(t) &= v_{x,0} \cos{\Omega t} + \frac{v_{x}}{2} 
    \sin{\Omega t},
\end{align}
which holds for the keplerian case.  

\begin{figure}[htbp]
    \centering
    \includegraphics[scale=0.6]{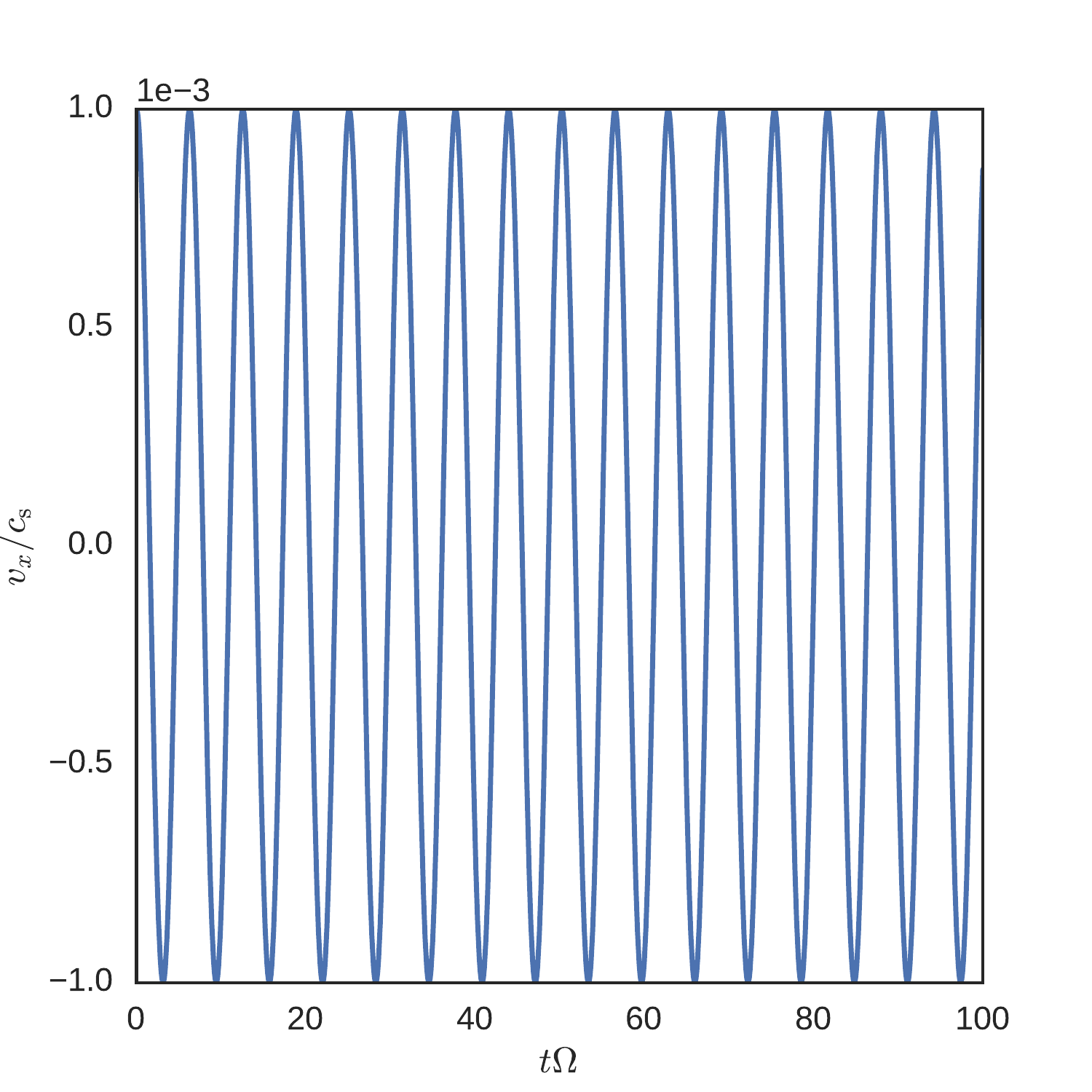}
    \caption{Velocity perturbation for the epicyclic motion test over \SI{100}{\Omega^{-1}}. There is no decline 
    or increase of the amplitude.}
    \label{fig:epicyclic_velocity}
\end{figure}

\cite{stone_implementation_2010} stress that the epicyclic energy
\begin{equation}
    E_{\mathrm{epi}} = \Sigma^2 \left( v_x^2 + \frac{2}{2+q} v_y^2 \right)
\end{equation}
should be conserved, which is achieved by a symmetric time integrator. Since \texttt{Fosite} is a purely explicit 
solver, we cannot conserve the epicyclic energy to round-off error. However, with the Dormand-Prince method 
\citep{dormand_family_1980}, we run the test over $\SI{1000}{\Omega^{-1}}$ and get a relative error of order $ 
10^{-5}$ for $E_{\mathrm{epi}}$. Additionally Dormand-Prince is of fifth order accuracy, compared to the 
Crank-Nicholson method, which is symmetric but only of second order. In fig. \ref{fig:epicyclic_velocity} the velocity 
evolution $v_x$ is shown over the first $\SI{100}{\Omega^{-1}}$.

\subsection{Linear Theory}
In order to test especially the gravitational solver (but also the boundaries and the fictitious forces) the equations 
of motion are linearized in $\Sigma = \Sigma_0 + \delta \Sigma \exp{\left( i \mathbf{k}(t) \cdot \mathbf{x} \right)}$ 
(similarly for other variables) and the evolution of the amplitude of a small density wave is examined. For the 
isothermal case this is derived in \cite{gammie_linear_1996}. Without magnetic fields and viscosity an ordinary 
differential equation 
\begin{align}\label{eq:lt_isothermal}
    \frac{\mathrm{d}^2 \delta \Sigma}{\mathrm{dt}^2} - \frac{\mathrm{d} \delta \Sigma}{\mathrm{d} t} 2 q & \Omega 
    \frac{k_x k_y}{|\mathbf{k}|^2} +  \nonumber \\ 
    \delta \Sigma  & \left( \Omega^2 -  2\pi G \Sigma_0 |\mathbf{k}| + c_{\mathrm{s}}^2 \mathbf{k}^2 + 
    -\tfrac{3}{2} 
    \Omega^2 \frac{k_y^2}{|\mathbf{k}|^2}\right)  = \nonumber  \\
    - 2 \Omega & \left( 1 - q \frac{k_y^2}{k^2}\right) \Sigma_0 \delta \xi.
\end{align}
is the result, where
\begin{equation}\label{eq:lt_isothermal_dxi}
    \delta \xi = \frac{1}{\Sigma_0}\left( \frac{\partial v_y'}{\partial x} - \frac{\partial v_x'}{\partial y} \right) - 
    \frac{(2-q) \Omega \delta \Sigma}{\Sigma_0^2}
\end{equation}
is the initial perturbation in the vorticity $\xi$. See also \cite{paardekooper_numerical_2012} for a thorough
derivation. In \cite{gammie_nonlinear_2001} the test is done including the energy equation, however we stay with the 
isothermal version to be comparable with \cite{paardekooper_numerical_2012}.


Similar to \cite{paardekooper_numerical_2012} we use eq.~\ref{eq:lt_isothermal} with an initial 
perturbation of $\delta \Sigma = \SI{5e-4}{\Sigma_0}$ and $\Sigma_0 = 1$. The wavenumbers are 
$k_{x,0} = -2(2\pi)$ and 
$k_{y,0} = 2 \pi$. 

\begin{figure}[htbp]
    \centering
    \includegraphics[scale=0.6]{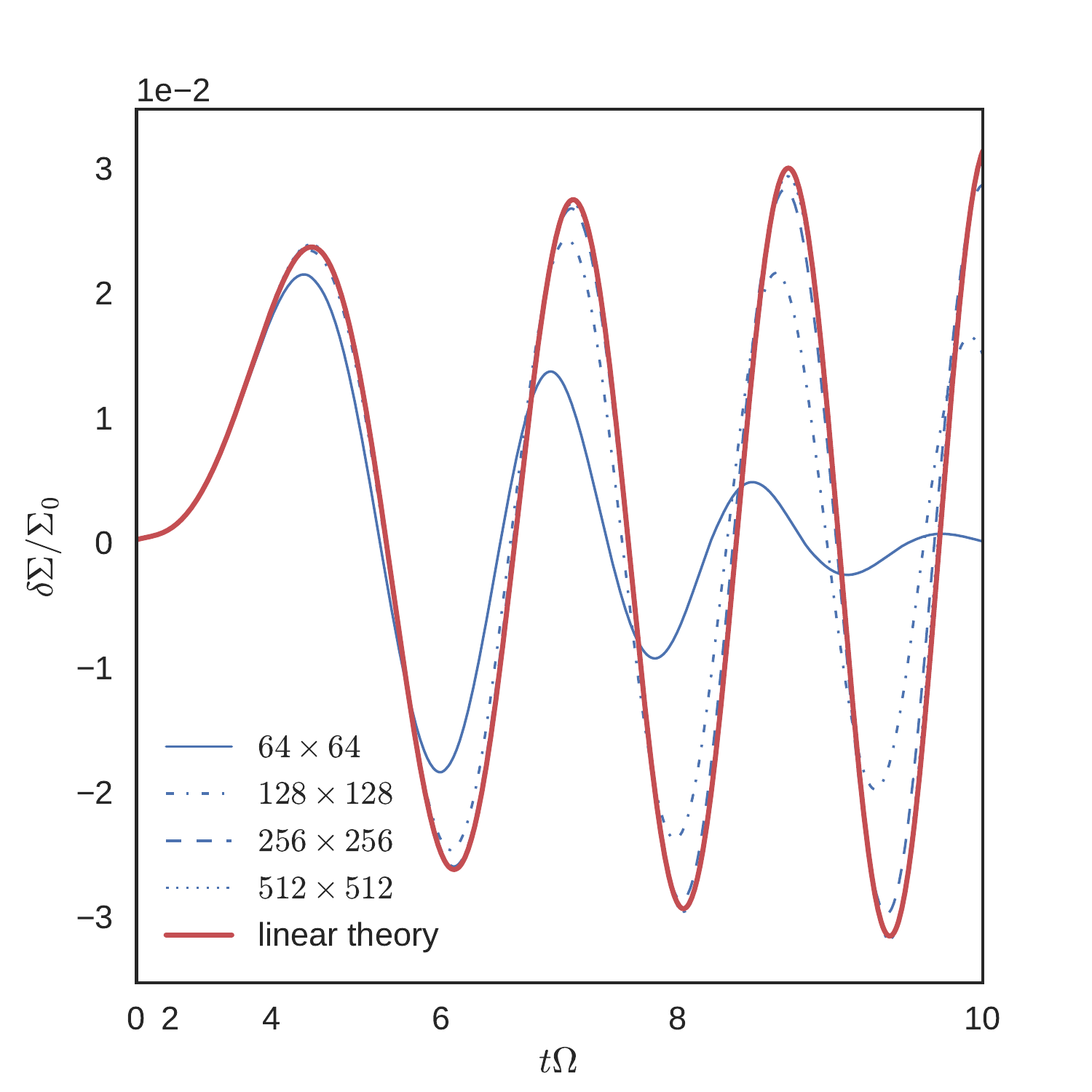}
    \caption{Density amplitude for the linear theory test with the \texttt{VANLEER} limiter. With increasinmg resolution, the numerical solution converges to the linear theory.}
    \label{fig:lt_convergence_vanleer}
\end{figure}
In fig. \ref{fig:lt_convergence_vanleer} we show convergence for the isothermal case with the \texttt{VANLEER} limiter. 
Typically, the first maximum is approximated well with all resolutions and limiters. At 
later stages the numerical viscosity plays a larger role with decreasing resolutions.

\begin{figure}[htbp]
    \centering
    \includegraphics[scale=0.6]{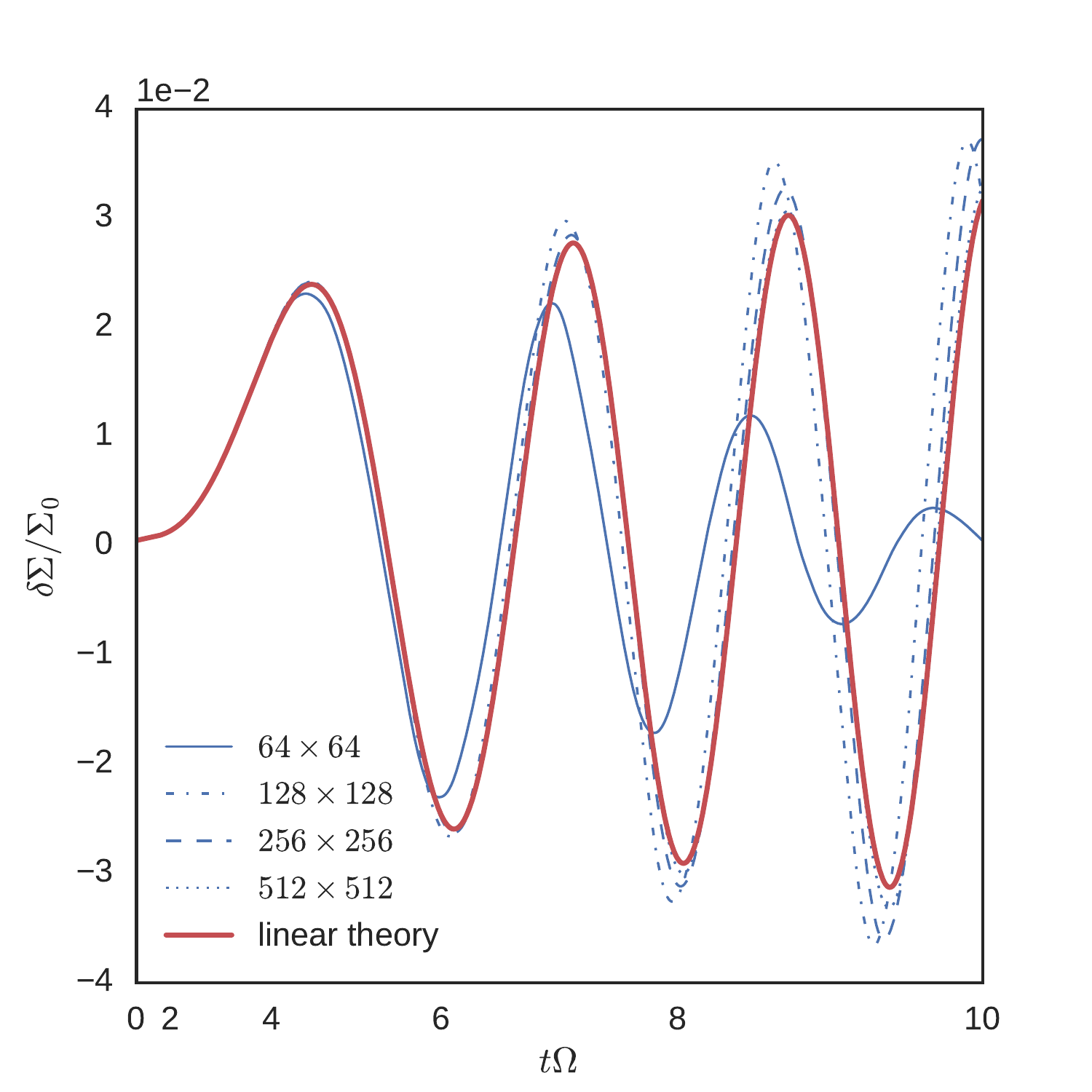}
    \caption{Density amplitude for the linear theory test with the \texttt{SUPERBEE} limiter. No clear convergence, but 
    rather an overestimation of the density amplitude is the case.}
    \label{fig:lt_noconvergence_superbee}
\end{figure}
Fig.~\ref{fig:lt_noconvergence_superbee} shows the same setup, but using the \texttt{SUPERBEE} limiter. There, we do 
not see convergence, but an overestimation of the analytical 
theory at higher resolution. The error is largest for a resolution of $N=128$ and then declines for higher resolutions. 
For $N=64$ however, the behavior is similar to the one in fig. \ref{fig:lt_convergence_vanleer}. Thus it seems that we 
have two opposing effects, one has a declining, viscous nature, which one should expect, since numerical dissipation 
usually declines with increasing resolutions. The other is strongly changing with the limiting function and it is only 
present or noticeable, if numerical dissipation is low enough, but leads to an excess in the density amplitude.

Since we expect that methods introduced to resolve shocks only add dissipation to the system, we would assume 
a lower amplitude over time. However, our test case inhibits no shocks, but shallow gradients. We therefore have strong 
indications that oversteepening caused by the \texttt{SUPERBEE} limiter is the reason for this effect 
\citep[see~e.g.][ch.~6; 
ch.~8]{leveque_finite_2002,hirsch_numerical_2007}.

\begin{figure}[htbp]
    \centering
    \includegraphics[scale=0.6]{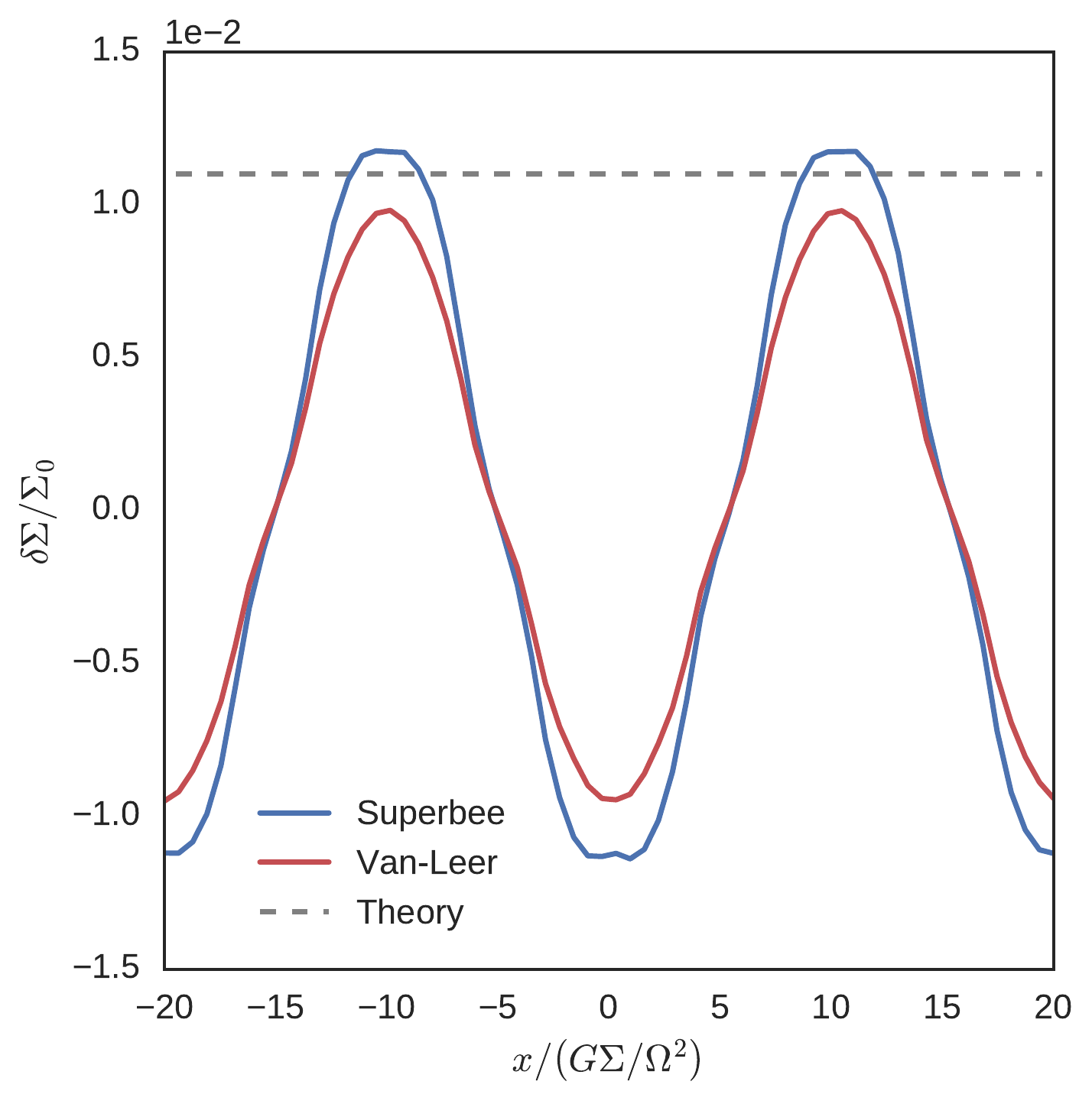}
    \caption{Cut through the density field along the $x$-axis at position $y=0$ and at time $t=\SI{520}{\Omega^{-1}}$. 
    The plateau of the \texttt{SUPERBEE} simulation is a clear indicator for oversteepening. The resolution is $N=64$.}
    \label{fig:oversteepening}
\end{figure}

It is always present at shallow regions but comes only into effect when resolution is high enough, because the inherent 
discretization error of the numerical scheme which leads to numerical dissipation decreases. Shallow, shock free 
regions are the expected environment where clumps form. To substantiate our thesis, we have a closer look at our 
calculation with gravity from fig.~\ref{fig:lt_convergence_vanleer} (\texttt{VANLEER}) and 
fig.~\ref{fig:lt_noconvergence_superbee} (\texttt{SUPERBEE}) at a resolution of $N=64$. In fig.~\ref{fig:oversteepening}
a cut through the backshifted density field (by $\Delta y = q \Omega x \mathbf{\hat{e}}_y$) along the $x$-axis at $y=0$ 
and time $t=520\Omega^{-1}$ is shown. This is shortly after the first transition through $\delta \Sigma=0$. Comparing 
the solutions, we see first that the \texttt{SUPERBEE} simulation is overestimating the theoretical prediction, but 
this is 
mainly because the numerical solution is forerunning the analytical one at low resolutions (see fig. 
\ref{fig:lt_noconvergence_superbee}). Second and most importantly the wave shows a clear plateau at the top of the 
waves. This is the common behavior and  a clear indicator\footnote{Together with steeper gradients, but the analytical 
solution gives us only the Fourier-component.} for an oversteepened function, 
which leads to considerable, non-physical deviations from the analytical solution.

\begin{figure}[htbp]
    \centering
    \includegraphics[scale=0.6]{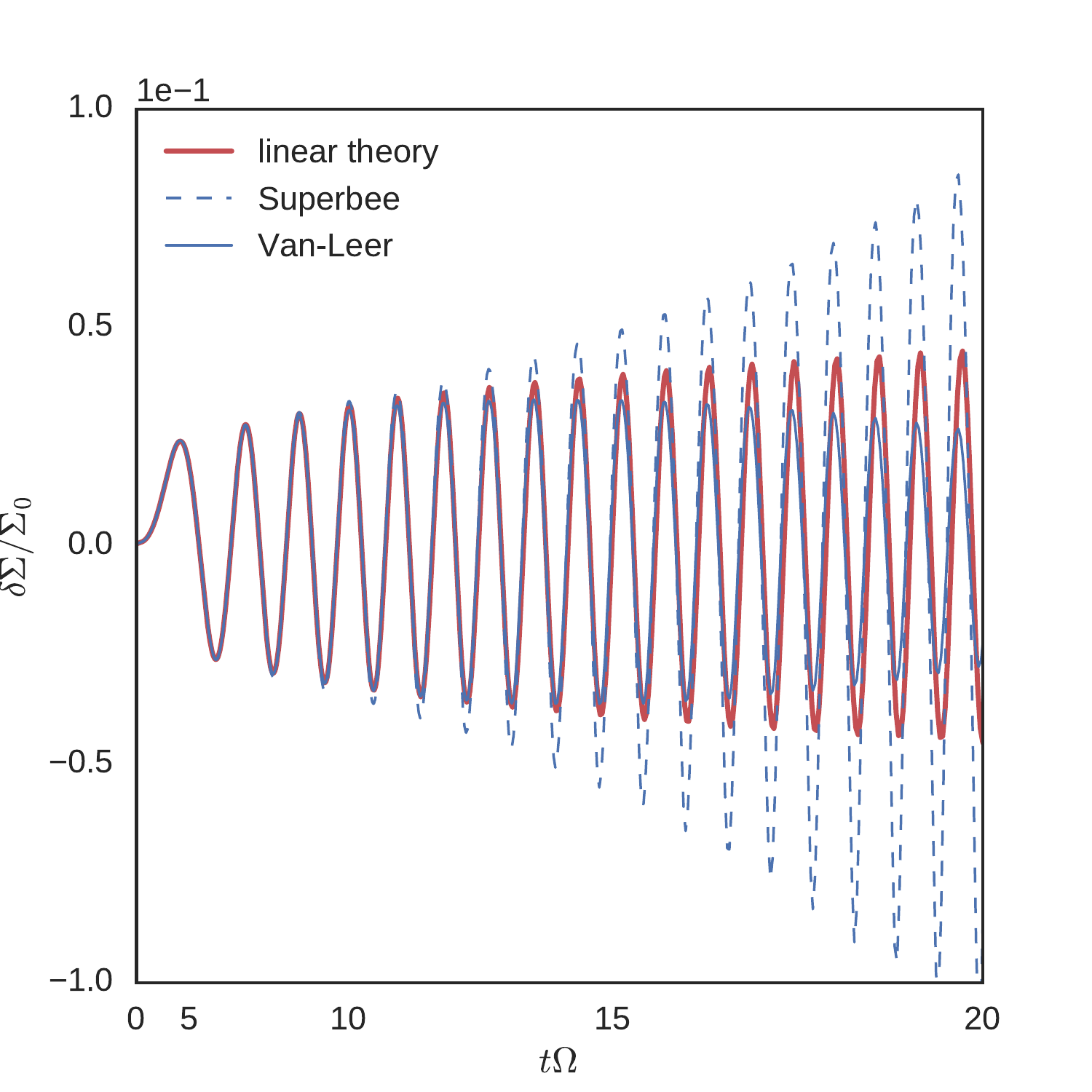}
    \caption{Linear theory test at a resolution of $512^2$. The future behavior strongly depends 
    on the limiting function and is qualitatively different.}
    \label{fig:lt_noconverge_long}
\end{figure}

\begin{figure}[htbp]
    \centering
    \includegraphics[scale=0.6]{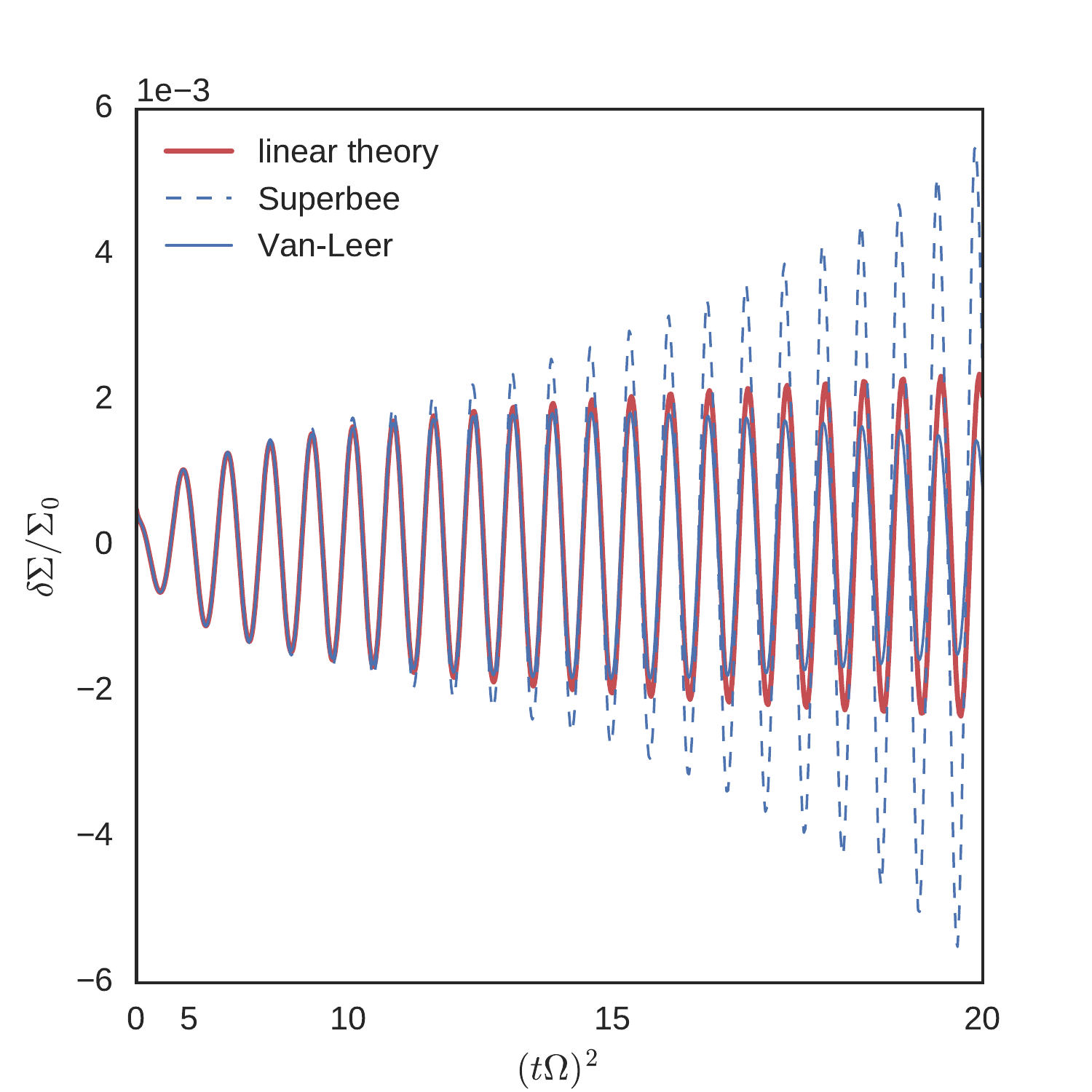}
    \caption{Linear theory test without gravity at a resolution of $512^2$. The effect does not dependent on the 
    self-gravitation but is purely numerical.}
    \label{fig:lt_nograv}
\end{figure}

We want to highlight under which circumstances the problem occurs in order to make predictions for a non-linear run. 
Therefore in fig.~\ref{fig:lt_noconverge_long}
the test is shown twice as long for the highest resolution of $512 \times 512$ with different limiters. Over these 
timescales it can be observed that the error amplifies increasingly and even 
if the first 
\SI{10}{\Omega^{-1}} seem to be converged, at later stages the numerical solution diverges. Without the gravitational 
module in fig. \ref{fig:lt_nograv}
the behavior remains the same, so gravitational forces are not the cause, but scale up initial errors. 

We also tested the influence of a different time integrator (strong-stability-preserving Runge-Kutta method of same 
order, \citealt{macdonald_constructing_2003}), a different Riemann solver (HLLC, 
\citealt{toro_restoration_1994,jung_resolving_2015}), with and without the advection splitting. We also carried out
calculations with the energy equation. There are minor differences for 
different values of $\gamma$, but in all tests we performed, the limiting function plays the dominant 
role for 
the diverging nature of the test. We cannot exclude that the effect does not appear for the \texttt{VANLEER} 
limiter 
at higher resolutions and over longer timescales. Lastly, we also crosschecked the simulations with additional 
limiters, namely \texttt{MINMOD} \citep{roe_characteristic-based_1986} and \texttt{OSPRE} 
\citep{waterson_unified_1995}, which also show convergence up to the resolution of $N=512$ (very slowly for 
\texttt{MINMOD}, 
however). 

There is an important difference to mention between our code and for example the one used by 
\cite{paardekooper_numerical_2012}. Our code is limiting the slopes in order to achieve a TVD-system 
\citep{illenseer_two-dimensional_2009} whereas \cite{paardekooper_numerical_2012} is using flux limiting 
functions for the same purpose. Having a look at comparing simulations of linear advection, it seems that 
slope-limiters are more prone to oversteepening than flux-limiters \cite[see][ch. 13.10]{toro_riemann_2009}. The effect 
exists however, in both cases.

\section{Nonlinear Results}\label{sec:results}
The non-linear results with the \texttt{VANLEER} and \texttt{SUPERBEE} limiting functions are tested and the 
fragmentation boundary is 
investigated. If they both only add dissipation to the system, they should converge at high enough resolutions to 
the same macroscopic behavior, since the artificial heating added to the system should be low enough to exclude this as 
the major driver of non-convergence, as, for instance, \cite{meru_convergence_2012} stated. 

\subsection{Standard run}
Similar to \cite{gammie_nonlinear_2001} we set $L = \SI{320}{G \Sigma / \Omega^2} \sim \SI{50}{H}$ with a 
resolution of $N=1024$ and $\beta=2$ or $\beta=10$, respectively, and in order to be comparable we also use 
$\gamma=2.0$. Until \SI{100}{\Omega^{-1}}, we can reproduce previous results with fragmentation for $\beta = 2$ and the 
gravito-turbulent state for $\beta = 10$. 

For the following calculations, we use an initial $\beta=20$ until $t=\SI{50}{\Omega^{-1}}$ and afterwards 
decrease $\beta$ for the next $\SI{50}{\Omega^{-1}}$ linearly to the desired value in order to avoid numerical 
problems 
from initializations \citep{paardekooper_numerical_2011}. 

With the \texttt{SUPERBEE} limiter and $\beta = 10$ fragmentation occurs at $t \simeq \SI{300}{\Omega^{-1}}$. We also can see 
the stochastic nature in this case. Therefore we ran the simulation with $\beta=15$ twice until $t = 
\SI{1100}{\Omega^{-1}}$, one simulation shows fragmentation, the other not. However, there is a difference to 
\cite{paardekooper_numerical_2012}: Over the whole simulation time, fragmentation occurs most probably up to $\beta 
\lesssim 12$. At the same time, with \texttt{VANLEER} limiter we see no fragmentation at all for 
$\beta > 2 $ at a resolution 
of $N=1024$. 

\subsection{The non-fragmenting case}
\begin{figure}[htbp]
    \centering
    \includegraphics[scale=0.45]{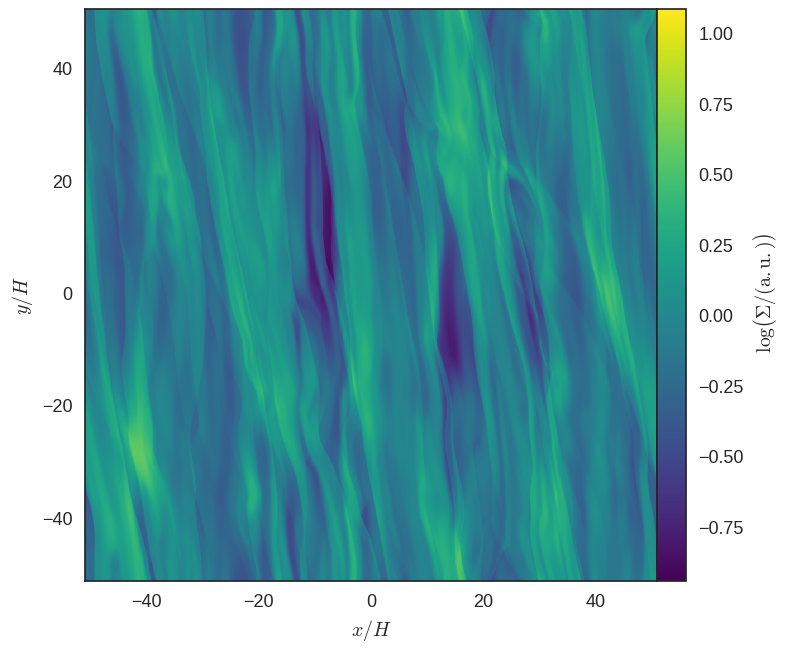}
    \caption{Standard run for \texttt{VANLEER} limiter at the end of the Simulation at $t=\SI{1100}{\Omega^{-1}}$.}
    \label{fig:stdrun_vanleer}
\end{figure}

\begin{figure}[htbp]
    \centering
    \includegraphics[scale=0.45]{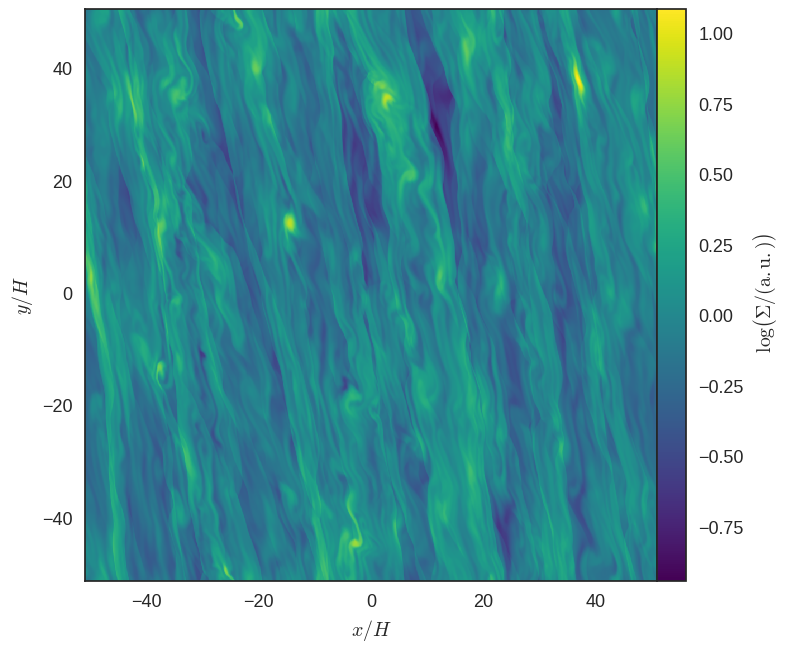}
    \caption{Standard run for \texttt{SUPERBEE} limiter at the end of the Simulation at $t=\SI{1100}{\Omega^{-1}}$.}
    \label{fig:stdrun_superbee}
\end{figure}

We compare the non-linear outcome for both limiters and a large value of $\beta$, where the gravito-turbulent state is 
present throughout the simulation time. In fig. \ref{fig:stdrun_vanleer} and fig. \ref{fig:stdrun_superbee} the 
macroscopic behavior can be seen at the end of the run with $t = \SI{1100}{\Omega^{-1}}$, for the \texttt{VANLEER} and 
\texttt{SUPERBEE} 
limiter, respectively. 
The simulation with \texttt{SUPERBEE} inhibits more structure and details. It shows small density enhancements, which 
did not 
lead to full fragmentation. These features are absent in the simulation with \texttt{VANLEER} limiter. Instead the gas 
is 
dominated by shocks running through the shearingsheet. Comparing our results with those of 
\citet[fig.~4]{gammie_nonlinear_2001} 
and \citet[fig.~1~-~left]{baehr_role_2015}, (with different cooling function) look much more like 
fig.~\ref{fig:stdrun_vanleer} at least in a qualitative way, whereas fig.~\ref{fig:stdrun_superbee} can be associated 
with results presented in \citet[fig.~6~-~top]{paardekooper_numerical_2012}.

\begin{figure}[htbp]
    \centering
    \includegraphics[scale=0.6]{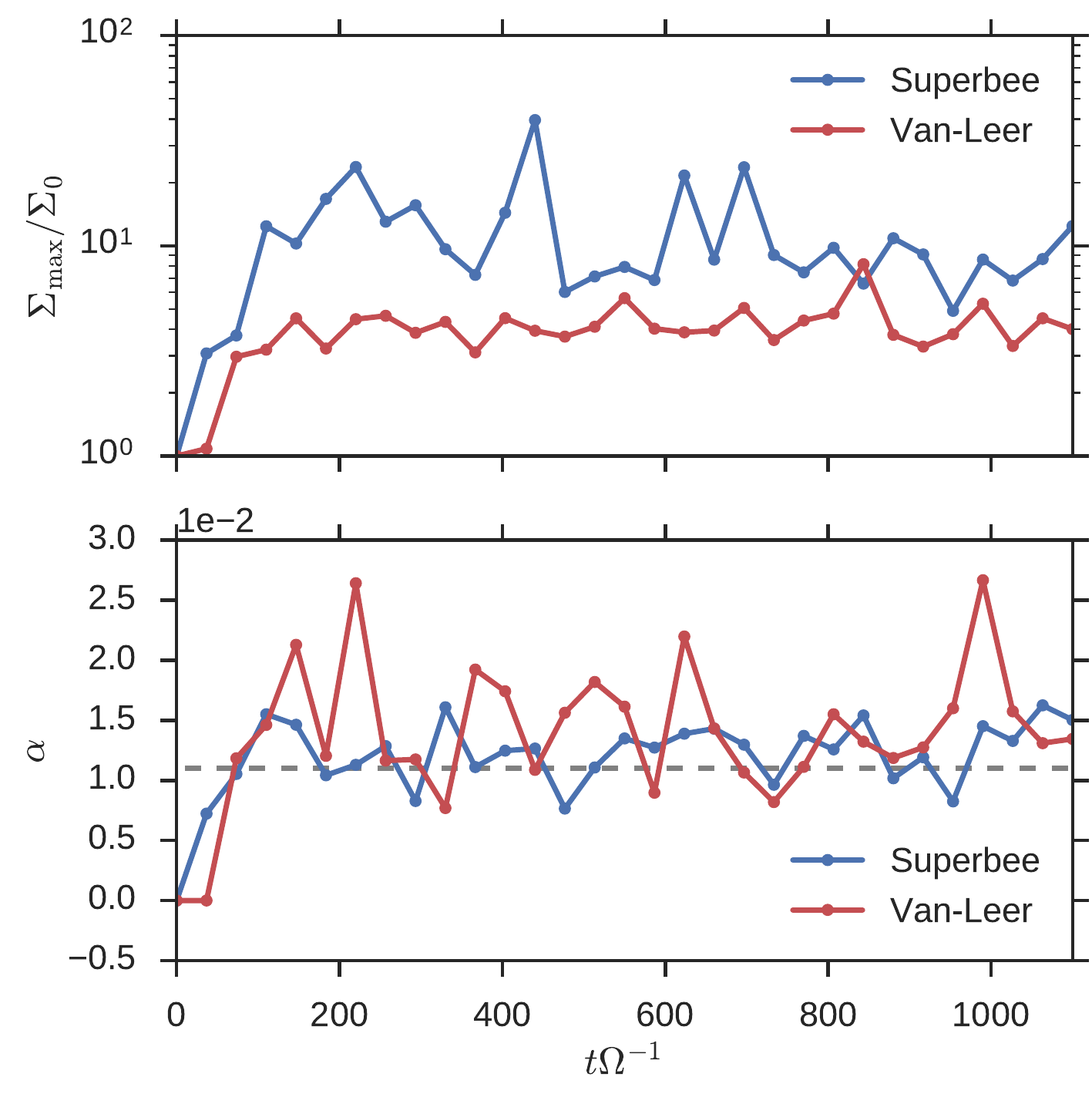}
    \caption{The maximum surface density $\Sigma_{\mathrm{max}}$ and the viscosity $\alpha$ for two different limiters 
    and $\beta=20$. The horizontal line is the theoretical value for $\alpha=0.011$.}
    \label{fig:SigmaMax_alpha}
\end{figure}

In fig.~\ref{fig:SigmaMax_alpha}
we show the maximum density $\Sigma_{\mathrm{max}}$ together with the measured value of 
$\alpha$. The simulation with the \texttt{SUPERBEE} limiter has generally a larger value for $\Sigma_{\mathrm{max}}$ 
and 
a larger amplitude of variations in $\Sigma$. If  $\beta$ were smaller these excesses may lead to fragmentation. At the same time, $\alpha$ 
yields similar values, with slightly better results with \texttt{SUPERBEE} limiter. The \texttt{VANLEER} limiter shows 
a larger variance 
around a value slightly above the theoretical value of $\alpha=0.011$ calculated by eq.~\ref{eq:alpha_theoretical}. 
This is probably the case, because the \texttt{VANLEER} limiter is more dissipative, leading to additional heating. 
It can, however, also be seen that this value is not very large and shows that numerical heating has a minor, but not 
fully 
neglectable impact on the simulations.
 
\begin{figure}[htbp]
\centering
\includegraphics[scale=0.6]{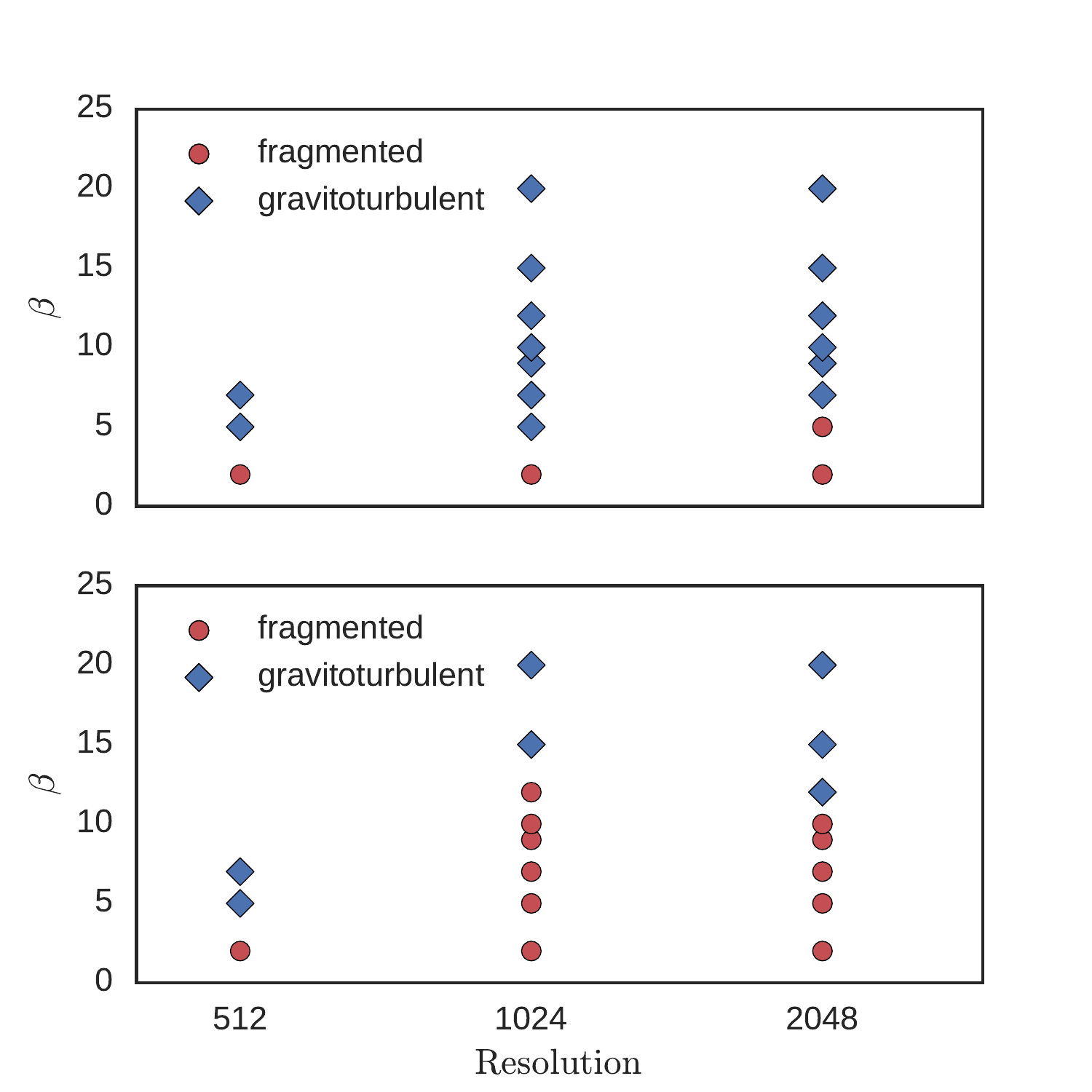}
\caption{Fragmentation plot for \texttt{VANLEER} limiter (top) and \texttt{SUPERBEE} limiter (bottom). The simulations 
where done for 
the resolution of $2048$ only up to \SI{300}{\Omega^{-1}}, compared to the other runs until \SI{1100}{\Omega^{-1}}. 
This is the reason why there is only fragmentation up to $\beta=10$ for the \texttt{SUPERBEE} limiter.}
\label{fig:frag_plot_vl_vs_sb.pdf}
\end{figure}

In fig.~\ref{fig:frag_plot_vl_vs_sb.pdf}
we see an overview of the fragmenting behavior. It shows fragmentation for $\beta<3$ throughout. For a resolution of 
$N=512$ both limiters have their fragmentation boundary directly at that value, very similar to 
\cite{gammie_nonlinear_2001}. At a larger resolution of $N=1024$ this behavior changes dramatically. The 
\texttt{VANLEER} 
limiter has its boundary value still at the same value for $\beta$ wheres the \texttt{SUPERBEE} limiter leads to 
fragmentation 
for at least up to $\beta=12$. At $N=2048$ \texttt{SUPERBEE} behaves very similar. For the \texttt{VANLEER} limiter the 
largest 
value with which we see fragmentation is $\beta=5$ at $N=2048$. It should be mentioned that we do not get the slow 
way of fragmentation in that case, but local collapse directly at $t=\SI{100}{\Omega^{-1}}$, where the final value of 
$\beta$ is reached. Interestingly, \cite{paardekooper_numerical_2012} obtains the same maximum result in 
$\beta_{\mathrm{crit}}$ for 
\texttt{MINMOD}, which generally is more dissipative. However, we cannot see any stochastic fragmentation for \texttt{VANLEER} 
limiter. 
It may still exists but then in a much tighter region around $\beta_{\mathrm{crit}} \sim 3-5$. Generally the 
simulations at a 
resolution of $N=2048$ where not performed over the whole \SI{1100}{\Omega^{-1}} but only up to \SI{300}{\Omega^{-1}}, 
which is most probably the reason that we do not get fragmentation for $\beta = 12$ and the \texttt{SUPERBEE} limiter. 

In order to check our results coherently in comparison to other studies we also made simulations with \texttt{MINMOD} limiter 
and 
with $\gamma=1.6$ ourself. \texttt{MINMOD} shows also fragmentation for $\beta<3$ at a resolution of $N=1024$ and no 
fragmentation for larger values. When choosing the smaller value for $\gamma$ we also see fragmentation 
for \texttt{VANLEER} limiter and $\beta=5$ at $N=1024$. This is consistent with \cite{rice_investigating_2005}, who 
state that 
lower a $\gamma$ leads to larger critical $\beta_{\mathrm{crit}}$.

\section{Discussion}\label{sec:discussion}
We showed that the choice of the most appropriate limiting function is crucial to the non-linear outcome of the simulations and 
cannot be seen just as a numerical helper adding more or less dissipation to the system. We showed this for the special 
case of Goduvov type schemes which make use of slope limiters in order to achieve total-variation-diminishing. 

\cite{rice_convergence_2014} showed that the cooling function needs to be adjusted in SPH simulations in order to 
converge. \cite{young_dependence_2015} however state that the gravitational smoothing leads to very different 
results in 2D simulations. Both publications change thereby physical quantities on small scales that are 
crucial for the fragmentation itself. Gravitational instabilities inherently depend on the cooling function and of 
course also on the gravitation itself. Thus, changing these quantities will always change the fragmentation 
behavior and could suppress any other physical or numerical effect on these scales. Still, we cannot exclude that these 
effects are the cause of fragmentation. \cite{young_dependence_2015} argue that smoothed 2D
simulations cannot resolve the quasi-static collapse within their description. In other contexts, however, like 
stochastic fragmentation, this poses no problem as 2D calculations by \citet{young_quantification_2016} demonstrate. 
The question of how strong this effect evolves in three dimensions and together with gravity is an open one.

Additionally we showed that oversteepening is also present without gravity and 
generally it is not affected by the number of dimensions, because of its pure numerical nature. The discussed 
numerical 
effect is neither dependent on gravitation, nor on the cooling function. We can also exclude 
that fragmentation for larger $\beta \sim 10$ is connected to numerical heating, since our results yield reasonable 
values for $\alpha$. It takes place in exactly the region where clumps form and the error occurs on timescales which 
are comparable to the orbital timescales. Additionally it can be backtracked to a well known effect called 
oversteepening and is testable with the linear theory test, at least in shearingsheet simulations. SPH codes introduce 
artificial stresses, which are always present, not only in shock regions, and some of them are especially introduced in 
order to avoid numerical fragmentation, because of, for instance, the testile instability \citep{monaghan_sph_2000}. This is why 
it cannot be excluded that SPH simulations also suffer from similar effects leading to an enhancement of 
density fluctuations which are in smooth regions scaled up by gravity.

This leads to an attractive property of oversteepening, as it may very well explain, why 
codes converge to different results or do not converge when $\beta>3$. The artificial viscosities or limiters, which are 
introduced in order to resolve shocks well, are handled completely differently by different kinds of codes and probably 
also handle smooth gradients completely different. 

Here, we showed that the effect exists for Godunov-type schemes. 
This is for example also the case in \cite{paardekooper_numerical_2012}. In that case, the limiting functions are 
applied to other physical quantities (fluxes instead of slopes), which may change but not remove the effect. Our 
simulations run in concordance with the estimated behavior, because the effect tends to be stronger for slope limiters 
in the literature \citep{hirsch_numerical_2007}, which we also see for our fragmentation behavior compared to 
\cite{paardekooper_numerical_2012}, both in the linear theory test and in the non-linear outcome. 

With the \texttt{SUPERBEE} limiter we can also see stochastic fragmentation like \cite{paardekooper_numerical_2012}. 
Running at a resolution $N=1024$ with $\beta=15$ over $t_{\mathrm{sim}}=\SI{1100}{\Omega^{-1}}$ twice, we observe 
fragmentation in one case only, but we did not investigate this issue in greater detail. 
With the other choice of the limiting function, we were not able to reproduce the stochastic features from 
\cite{paardekooper_numerical_2012}. They may indeed exist, but then in a much lower range then previously expected 
(around $\beta_{\mathrm{crit}} \lesssim 5$). This is also consistent with the results of 
\cite{paardekooper_numerical_2012} himself, 
since he got $\beta_{\mathrm{crit}} \lesssim 5$ for the very dissipative limiter \texttt{MINMOD}, where most probably 
no oversteepening is 
present. With the \texttt{VANLEER} limiter, which is less dissipative, but also does not show oversteepening within the 
test, we get the same critical value for $\beta_{\mathrm{crit}}$. Furthermore, independent SPH simulations by 
\cite{young_quantification_2016} show that the effect is not as strong as expected before.

Another point important to mention is that the statement of \cite{young_dependence_2015} that there is no 
dependency on $\gamma$ in two-dimensional simulations is not coherent with our results. We can for 
example see fragmentation for $\beta=5$ and the \texttt{VANLEER} limiter, when choosing $\gamma=1.6$ and a 
resolution of $N=1024$. This is in concurrency with the results of \cite{rice_investigating_2005}, where it was stated 
that smaller $\gamma$ lead to larger critical $\beta_{\mathrm{crit}}$. Additionally, heat capacity ratios are often 
compared between 
two and three dimensions equivalently and although they might behave similarly, they are not the same 
\citep{gammie_nonlinear_2001}. 

%
%

\section{Conclusions}\label{sec:conclusion}
In this paper, we investigated the impact of the limiting function on the fragmentation boundary in self-gravitating 
shearing-sheet simulations. By comparing the numerical outcome with linear theory, we showed that there is strong 
evidence that a unfortunate choice of the limiter can lead to an overestimation in the surface density in regions with 
shallow density gradients, which makes fragmentation easier for larger $\beta$. We accomplish this by using 
two-dimensional shearingsheet simulations and looking at deviations from to analytical results for linear theory. The 
behavior can be ascribed to a well known numerical effect called oversteepening, which occurs in regions with shallow 
density gradients. It is only observable if the resolution is high and the corresponding numerical diffusion low 
enough. Additionally we can exclude gravity as the driving force of the effect, but it is scaling the small initial 
errors up. For the limiting functions where oversteepening sets in place in the test, we see very strong differences in 
the non-linear outcome. In these calculations we need much smaller critical $\beta_{\mathrm{crit}}$, because of 
abundant density 
excesses in the field. We also cannot reproduce stochastic fragmentation with a well chosen limiter, however 
this might take place in a much smaller range for values of $\beta$.

\begin{acknowledgements}
We thank an anonymous referee for very helpful comments on an earlier version of this paper.
\end{acknowledgements}

\bibliographystyle{aa}
\bibliography{references}
\end{document}